\documentclass[pra,twocolumn,superscriptaddress,preprintnumbers,amsmath,amssymb,nofootinbib,floatfix]{revtex4-2}
\usepackage[utf8]{inputenc}
\usepackage{amsmath}
\usepackage{comment}
\usepackage{siunitx}
\usepackage{dcolumn}
\usepackage{braket}
\usepackage{chemformula}
\usepackage{graphicx,bm}
\usepackage{float}
\usepackage{subfigure}
\usepackage{amsmath}
\usepackage{amsfonts}
\usepackage{relsize}

\usepackage{tikz}
\usepackage{inputenc}
\usepackage{amsthm, amssymb,bbm,bm}
\usepackage{hyperref}
\usepackage{siunitx}
\usepackage{tabularx}
\usepackage{multirow}

\begin{document}

\title{Theory of photonic crystal polaritons in periodically patterned multilayer waveguides}
\author{Simone Zanotti}
\affiliation{Dipartimento di Fisica, Universit\`{a} di Pavia, via Bassi 6, I-27100 Pavia, Italy}
\author{Hai Son Nguyen}
\affiliation{Univ Lyon, Ecole Centrale de Lyon, CNRS, INSA Lyon, Universit\'e  Claude Bernard Lyon 1, CPE Lyon, CNRS, INL, UMR5270, Ecully 69130, France}
\affiliation{Institut Universitaire de France (IUF), Paris, France }
\author{Momchil Minkov}
\affiliation{Flexcompute Inc., 130 Trapelo Rd., Belmont, Massachusetts 02478, USA}
\author{Lucio C. Andreani}
\affiliation{Dipartimento di Fisica, Universit\`{a} di Pavia, via Bassi 6, I-27100 Pavia, Italy}
\author{Dario Gerace}
\affiliation{Dipartimento di Fisica, Universit\`{a} di Pavia, via Bassi 6, I-27100 Pavia, Italy}

\begin{abstract}
We present a formalism for studying the radiation-matter interaction in multilayered dielectric structures with active semiconductor quantum wells patterned with an in-plane periodic lattice. The theory is based on the diagonalization of the generalized Hopfield matrix, and it includes loss channels in a non-Hermitian formulation. Hybrid elementary excitations named photonic crystal polaritons arise in these systems, whose detailed dispersion and loss characteristics are shown to depend on material composition as well as on symmetry properties of the lattice. We show the generality of the approach by calculating polariton dispersions in very diverse material platforms, such as multilayered perovskite-based lattices or inorganic semiconductor heterostructures.  As an application of the method, we show how to engineer lossless polariton modes through excitonic coupling to bound states in the continuum at either zero or finite in-plane wavevector, and discuss their topological properties. Detailed comparison with a semiclassical approach based on the scattering matrix method is provided, which allows to interpret the optical spectra in terms of polarization-dependent excitation of the different polariton branches. This work introduces an efficient and invaluably versatile numerical approach to engineer photonic crystal polaritons, with potential applications ranging from low-threshold lasers to symmetry-protected propagating modes of hybrid radiation-matter states.
\end{abstract}



\maketitle

\section{Introduction}


Non-perturbative radiation-matter coupling between dipole-active material excitations and eigenmodes of the electromagnetic field leads to the concept of polaritons, i.e., hybrid excitations of mixed nature involving at least two fields with different characteristics \cite{art:Basov21}. Microcavity exciton-polaritons, in particular, arising from the strong coupling between bound electron-hole pairs in semiconductor quantum wells (QW excitons) and planar microcavity photons, have been shown to behave as low-mass elementary excitations with bosonic-like statistics \cite{art:Sanvitto2016}, thus allowing to unravel phenomena that are typically displayed in weakly interacting bosonic gases \cite{art:Carusotto-Ciuti2013}, such as Bose-Einstein condensation \cite{art:Kasprzak2006} and superfluidity \cite{art:Amo2009}, with potential applications in all-optical devices such as low-threshold lasing \cite{art:Azzini11}, all-optical switching \cite{art:Ballarini2013} and routing \cite{art:Nguyen13,art:Marsault2015}, and more recently  quantum technologies \cite{art:Davide22}. The concept of exciton-polaritons has then been extended to excitons coupled to periodically patterned metasurfaces with sub-wavelength pitch, also named photonic crystal polaritons \cite{art:Dario_Pol}. These excitations have been shown to arise in very diverse materials and platforms, ranging from inorganic semiconductor heterostructures \cite{art:Bajoni09,art:Whittaker21,art:Ardizzone2022}, to two-dimensional materials (TDM) deposited on a patterned metasurface \cite{art:Koshelev18,art:monolayer,art:TMD,art:Arka20} or layered perovskites within a passive backbone \cite{art:Ishihara98,art:Ha_My,art:Koreani21}. As it will be shown in the following, in all these different scenarios the physics of photonic crystal polaritons can essentially be described within a common theoretical background. 

While polaritons can be understood as semiclassical excitations of the electromagnetic field in a polarizable medium \cite{Andreani_Book_2014}, a full quantum-mechanical theory of polaritons was originally proposed by Hopfield \cite{art:Hopfield} and Agranovich \cite{Agranovich1960} in two seminal papers. In particular, the Hopfield formulation is based on a transformation of operators and can be extended to include imaginary parts of the oscillator energies, thus making the theory a non-hermitian one. While a generalized formulation of the Hopfield method has been reported in the literature \cite{Ciuti2005,Hagenmuller2010}, a Hopfield-based quantum theory accounting for an arbitrary multilayered structures with an in-plane pattern is still lacking. In this work, we extend a previously formulated theory of radiation-matter interaction in fully etched thin photonic crystal slabs containing a single active layer \cite{art:Dario_Pol} to describe an arbitrarily etched multilayered structure. In particular, we consider a guided mode expansion (GME) basis for the approximate photonic modes that are the closed system solution of a multilayered planar waveguide with partially etched core \cite{art:GME,art:GME_Momchil}, in which losses are calculated within a perturbative approach by coupling to the continuum of radiative modes in the light cone \cite{art:GME} to obtain the eigenvalues imaginary parts. Then we formulate a generalized Hopflied Hamiltonian accounting for radiation-matter coupling to an arbitrary number of active layers characterized by an excitonic response. We generalize previous treatments by including the polarization response of the excitonic transition. 

As applications of this non-Hermitian theory, we show that the scaling of the polariton Rabi splitting deviates from the expected $\sqrt{K}$ dependence as a function of the number of active layers ($K$), due to the guided mode profile that weights the light-matter coupling on the waveguide thickness. Then, we show that the polariton dispersion in square lattices of perovskite metasurfaces can be obtained, and it compares favorably well with either rigorous coupled wave analysis simulations. 
Finally, we further demonstrate the generality and versatility of the approach by analysing the effect of strong light-matter coupling between QW excitons and bound-states in the continuum (BIC) supported in partially etched triangular lattice photonic crystal slab. We perform a systematic analysis of the topological properties of such polariton BICs, relating in particular the topological charge to the polarization singularity in reciprocal space. These studies acquire particular significance in light of recent experiments unravelling the topological nature of photonic crystal polaritons \cite{Kravtsov2020,art:Ha_My,art:Koreani21,art:Ardizzone2022}. Finally, we clarify the role of non-radiative versus radiative loss contributions in the BIC optical response.

\section{Theory }

\begin{figure}[t]
\centering
    \includegraphics[width=\linewidth]{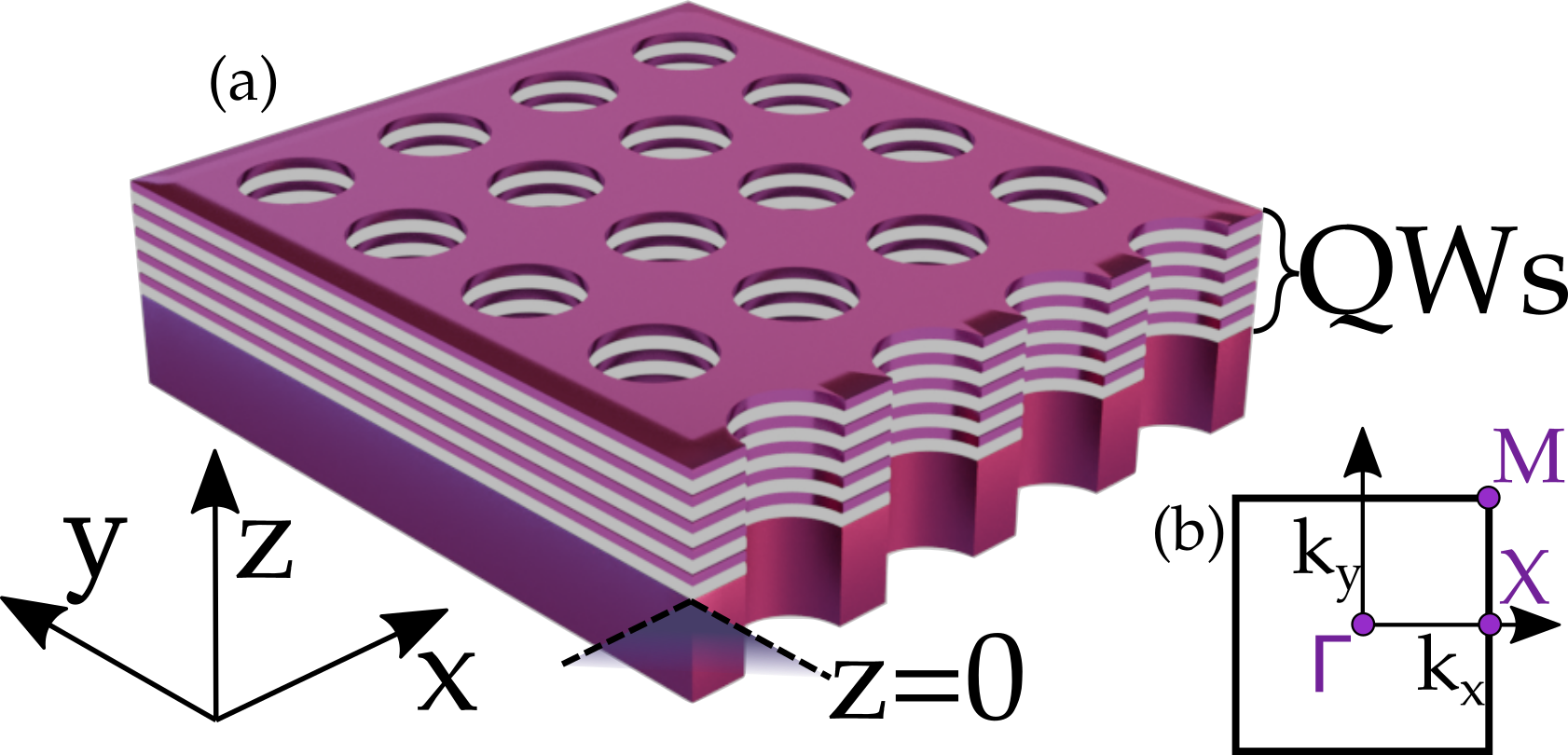}
    \caption{(Colour online) (a) Photonic crystal slab with quantum wells embedded from the centre of the slab in $z=0$ to the top. (b) Brillouin zone with high-symmetry points for a square lattice.}
    \label{fig:sketch_mqws}
\end{figure}

The class of systems we consider in this work are schematically represented in Fig.~\ref{fig:sketch_mqws}, where the active layers (i.e., QWs containing bound electron-hole pairs) are coupled to a periodic photonic landscape. In the following, we summarize the basic steps to formulate a Hopfield matrix description of this problem.

\subsection{Photonic modes: the Guided-Mode Expansion}\label{subsec:GME}

We start with the formalism to numerically calculate the non-Hermitian photonic modes by the guided-mode expansion (GME) method \cite{art:GME}\cite{art:GME_Momchil}. This method is particularly efficient in calculating the (approximated) photonic dispersion (real part of the eigenvalues) of layered structures, with an arbitrary in-plane periodicity, as well as their imaginary part (related to out-of-plane scattering losses) within perturbation theory. Here we give a brief summary of the method for completeness, and to fix the notation for the following.

The GME method assumes the presence of linear, isotropic, non-dispersive, lossless, and non magnetic materials in absence of free charges. Under such hypotheses,  Maxwell equations can be written as an eigenvalue problem for the harmonic magnetic field, $\mathbf{H}(\mathbf{r},t)=\mathbf{H}(\mathbf{r})e^{-i\omega t}$:
\begin{equation}
    \hat{\Theta}\mathbf{H}(\mathbf{r})=\frac{\omega^2}{c^2}\mathbf{H}(\mathbf{r}),\quad \hat{\Theta}\doteq\nabla\times\frac{1}{\varepsilon(\mathbf{r})}\times,
\end{equation}
where $\varepsilon(\mathbf{r})$ is the space-dependent relative permittivity. The electric field $\mathbf{E}(\mathbf{r},t)=\mathbf{E}(\mathbf{r})e^{-i\omega t}$ can be recovered from the magnetic one:
\begin{equation}
    \mathbf{E}(\mathbf{r}) = \frac{i}{\omega \varepsilon_0 \varepsilon(\mathbf{r})}\nabla\times \mathbf{H}(\mathbf{r}).
\end{equation}
In the GME method, the magnetic field is expanded in a basis of orthonormal modes:
\begin{equation}
    \mathbf{H}(\mathbf{r})=\sum_\mu c_\mu\mathbf{H}_\mu(\mathbf{r}),\quad \int_{\text{unit cell}}  \mathbf{H}_\mu(\mathbf{r})\cdot\mathbf{H}_\nu(\mathbf{r})d\mathbf{r} =\delta_{\mu,\nu},
\end{equation}
leading to the linear eigenvalue problem:
\begin{eqnarray}\label{eq:GME_mat_el}
    &&\sum_\nu \mathcal{H}_{\mu\nu}c_\nu =\frac{\omega^2}{c^2}c_\mu,\nonumber\\
    && \mathcal{H}_{\mu,\nu}= \int\frac{1}{\varepsilon(\mathbf{r})}\left(\nabla\times \mathbf{H}_\nu^*(\mathbf{r}) \right)\cdot\left(\nabla\times \mathbf{H}_\mu(\mathbf{r}) \right)d\mathbf{r}.
\end{eqnarray}
In order to evaluate the matrix elements in \eqref{eq:GME_mat_el}, the magnetic field is expanded for each in-plane wavevector $\mathbf{k}$ in the suitable basis of the guided modes:
\begin{equation}
    \mathbf{H}_\mathbf{k}(\mathbf{r})=\sum_{\mathbf{G},\alpha} c(\mathbf{k}+\mathbf{G},\alpha)\mathbf{H}^{\text{guided}}_{\mathbf{k}+\mathbf{G},\alpha}(\mathbf{r}),
\end{equation}
where $\mathbf{G}$ is a reciprocal lattice vector, $\alpha$ is index of the guided mode which can be explicitly given as:
\begin{equation}
    \mathbf{H}^{\text{guided}}_{\mathbf{k}+\mathbf{G},\alpha}(\mathbf{r})= \mathbf{h}_{\mathbf{k}+\mathbf{G},\alpha}(z)e^{i(\mathbf{k}+\mathbf{G})\cdot\rho},
\end{equation}
$\rho$ being the in-plane position vector $(x,y)$, and $\mathbf{h}_{\mathbf{k}+\mathbf{G},\alpha}(z)$ the profile of the guided mode.

This procedure leads to purely real frequencies of the photonic modes, $\omega_{\mathbf{k}n}$. The radiative losses of leaky modes lying above the light line are calculated with the photonic analogue of the Fermi golden rule, yielding the (wave vector-dependent) imaginary part of each eigenvalue as:
\begin{equation}\label{eq:im-part}
    \Im\left(\frac{\omega^2_\mathbf{k}}{c^2}\right)=-\sum_{\mathbf{G},p}\:\sum_{o=l,u}|\mathcal{H}_{\mathbf{k}r}|^2\rho_0\left(\mathbf{k}+\mathbf{G},\omega_{1}\right),
\end{equation}
where $\rho_0$ is the 1D density of photonic states, $\mathcal{H}_{\mathbf{k}r}$ is the coupling matrix element defined as:
\begin{equation}
    \mathcal{H}_{\mathbf{k}r}=\int\frac{1}{\varepsilon(\mathbf{r})}\left(\nabla\times\mathbf{H}^*_{\mathbf{k} }(\mathbf{r})  \right)\cdot\left(\nabla\times\mathbf{H}^{\text{rad}}_{\mathbf{k}+\mathbf{G},p,o }(\mathbf{r})  \right)d\mathbf{r},
\end{equation}
$p$ represents the polarisation of the radiative modes (TE,TM), while the summation over $o$ takes into account the leakage through the upper ($u$) and lower ($l$) cladding. 

Starting from the second-quantised formalism we can write the photonic Hamiltonian as:
\begin{equation}\label{eq:ham_pho}
    H_{\text{ph}}=\sum_{\mathbf{k},n}\hbar\omega_{\mathbf{k}n}\left(
    \hat{a}_{\mathbf{k}n}^\dagger\hat{a}_{\mathbf{k}n} +\frac{1}{2}\right),
\end{equation}
where $\omega_{\mathbf{k}n}$ is the (real) frequency of the n-th band at given wave vector $\mathbf{k}$ calculated with GME, and $\hat{a}_{\mathbf{k}n}$ is the annihilation operator which follows bosonic commutation rules:
\begin{equation}
    \left[\hat{a}_{\mathbf{k}n},\hat{a}^\dagger_{\mathbf{k}'n'}\right]=\delta_{\mathbf{k},\mathbf{k}'}\delta_{n,n'},\quad \left[\hat{a}_{\mathbf{k}n},\hat{a}_{\mathbf{k}'n'}\right]=\left[\hat{a}^\dagger_{\mathbf{k}n},\hat{a}^\dagger_{\mathbf{k}'n'}\right]=0
\end{equation}
When generalizing to the non-Hermitian Hopfield matrix, the photonic eigenvalues will be taken as the complex eigenmodes $\tilde{\omega}_{\mathbf{k}n}=\omega_{\mathbf{k}n} - i \gamma_{\mathbf{k}n}$ with $\gamma_{\mathbf{k}n}=-\Im\left[\sqrt{\Re\left(\omega^2_{\mathbf{k}n}\right)+i\Im\left(\omega^2_{\mathbf{k}n}\right)}\right]> 0$, in which $\Im\left(\omega^2_{\mathbf{k}n}\right) < 0$ is obtained from Eq.~(\ref{eq:im-part}), see Appendix.

\subsection{Exciton envelope function in a periodic potential}\label{subsec:sch}
Excitons are collective excitations formed by an electron-hole bound state.
We will treat excitons as a non-interacting two-dimensional gas of bosonic quasi-particles.
This assumption is justified in the low-density regime \cite{art:Carusotto-Ciuti2013}, where the excitonic density, $n_{\text{ex}}$, is much smaller than the saturation density, $n_{\text{sat}}\approx1/(2\pi a_{\text{2D}})$. Excitons in two-dimensionally confined semiconductors and insulators can be treated within the envelope function approximation, yielding the two-body function of electron and hole \cite{art:andreani_cilyndrical}.

In the present case of multilayered systems, we start from the factorisation of the excitonic envelope function written as:
\begin{equation}
    F(\mathbf{r}_e,\mathbf{r}_h)= F_\mathbf{k}(\mathbf{R}_{||})f(\rho,z_e,z_h),
\end{equation}
where $\mathbf{R}_{||}$ and $\rho$ are the in-plane exciton center of mass and bound electron-hole pair relative coordinates, respectively.
The in-plane dynamics of the excitons can be calculated by solving the Schr\"{o}dinger equation for the envelope function:
\begin{equation}
    \left[ -\frac{\hbar^2\nabla^2}{2M}+V\left(\mathbf{R}_{||}\right)\right]F_{\mathbf{k}}\left(\mathbf{R}_{||}\right)=\left(E_\mathbf{k}-E_{\text{ex}}\right)F_{\mathbf{k}}\left(\mathbf{R}_{||}\right),
\end{equation}
with $E_{\text{ex}}$ being the bare exciton energy of the active material and $M=m^*_e+m^*_h$ the exciton mass; $V\left(\mathbf{R}_{||}\right)$ is the in-plane potential felt by excitons, see Sec. \ref{sec:app} for a number of examples.
Similarly to the plane wave expansion of the magnetic field in the GME method, the envelope wave function can be expanded on the same set of plane wave modes, as:
\begin{equation}
    F_{\mathbf{k}}\left(\mathbf{R}_{||}\right)=\sum_\mathbf{G}F\left(\mathbf{k}+\mathbf{G}\right)e^{i\left(\mathbf{k}+\mathbf{G}\right)\cdot\mathbf{R}_{||}},
\end{equation}
leading to the Schr\"{o}dinger-type wave equation in reciprocal space, which is numerically easy to solve:
\begin{eqnarray}\label{eq:sch_rec}
    &&\sum_{\mathbf{G}'}\left[ \frac{\hbar^2|\mathbf{k}+\mathbf{G}|^2}{2M}\delta_{\mathbf{G},\mathbf{G}'}+V\left( \mathbf{G}-\mathbf{G}'\right) \right]F\left(\mathbf{k}+\mathbf{G}'\right)\nonumber\\
    & &=\left(E_\mathbf{k}-E_{\text{ex}}\right)F\left(\mathbf{k}+\mathbf{G}\right),
\end{eqnarray}
where $V\left( \mathbf{G}-\mathbf{G}'\right)$ is the Fourier component of the in-plane potential. Following the diagonalization of the wave equation, we can write a quantised excitonic Hamiltonian by introducing the annihilation operator $\hat{b}_{\mathbf{k},\nu}$ for excitonic quasi-particles:
\begin{equation}\label{eq:ham_ex}
    H_{\text{ex}}=\sum_{\mathbf{k},\nu}E_{\mathbf{k}\nu}\hat{b}^\dagger_{\mathbf{k}\nu}\hat{b}_{\mathbf{k}\nu}.
\end{equation}

In view of formulating the generalized Hopfield matrix, the 
intrinsic excitonic losses (either radiative or non-radiative) can be added \textit{a posteriori} through an imaginary term of the energies, $\gamma_{\text{ex}}=\Gamma_{\text{ex}}/2$, where $\Gamma_{\text{ex}}$ is intended as the resonance linewidth (i.e., the full width at half maximum), i.e., by defining $\tilde{E}_{\mathbf{k}\nu}=E_{\mathbf{k}\nu}-i\gamma_{\text{ex}}$ (see Appendix).

\subsection{Generalized Hopfield theory of exciton-photon coupling }
Now that we have described the purely photonic and excitonic pictures in second quantization, see Eqs.~\eqref{eq:ham_pho} and \eqref{eq:ham_ex}, we can take into account
their mutual interaction. Radiation-matter interaction can be described by the minimal coupling Hamiltonian written in terms of the vector potential, $\mathbf{A}$, and electron position ($\mathbf{r}$) and momentum ($\mathbf{p}$) within a generalised Coulomb gauge, where $\nabla\cdot(\varepsilon\mathbf{A})=0$:
\begin{align}\label{eq:ham_min}
\begin{split}
    H_I = -\frac{e}{4m_0}\sqrt{ \frac{1}{\pi\varepsilon_0 c^2}}
    \sum_{l=1}^L\{\mathbf{A}(\mathbf{r}_l)\cdot\mathbf{p}_l+\mathbf{p}_l\cdot\mathbf{A}(\mathbf{r}_l)\}\\
    +\frac{e^2}{8\pi m_0\varepsilon_0 c^2}\sum_{l=1}^L|\mathbf{A}(\mathbf{r}_l)|^2= H_I^{(1)}+H_I^{(2)}.
\end{split}
\end{align}
The two parts of the Hamiltonian \eqref{eq:ham_min} can be rewritten in terms of the photonic and excitonic destruction (creation) operators, $\hat{a}$ ($\hat{a}^{\dagger}$) and $\hat{b}$ ($\hat{b}^{\dagger}$), respectively \cite{art:Dario_Pol}:
\begin{eqnarray}
     &&H_I^{(1)} = i\mkern-18mu\;\sum_{\mathbf{k},n,\nu,\sigma,j}\!\!\!\!C_{\mathbf{k}n\nu\sigma j}
    (\hat{a}_{\mathbf{k}n}+\hat{a}_{-\mathbf{k}n}^\dagger)(\hat{b}_{-\mathbf{k}\nu\sigma j}-\hat{b}_{\mathbf{k}\nu\sigma j}^\dagger)\\
    && H_I^{(2)}=\mkern-18mu\!\!\sum_{\mathbf{k},nn',\nu,\sigma,j }\!\!\!\!\!D_{\mathbf{k}nn'\nu\sigma j }
    (\hat{a}_{-\mathbf{k}n}+\hat{a}_{\mathbf{k}n}^\dagger)(\hat{a}_{\mathbf{k}n'}+\hat{a}_{-\mathbf{k}n'}^\dagger),
\end{eqnarray}
where $n\in\left[1,N\right]$ is the photonic band index , $\nu\in\left[1,M\right]$ denotes the excitonic band index. We also notice that the present formulation generalizes the previous one \cite{art:Dario_Pol}, since we now also consider indices $j\in\left[1,K\right]$ corresponding to the active layer within the vertical heterostructure, and $\sigma=1,2,3$ corresponding to the exciton polarization (along $x$, $y$, or $z$).

The coupling terms can be formally calculated as $D_{\mathbf{k}nn'\nu\sigma j } = C^*_{\mathbf{k}n\nu\sigma j} C_{\mathbf{k}n'\nu\sigma j}/E_{\mathbf{k},\nu,\sigma,j}$, where $C_{\mathbf{k}n\nu\sigma j}$ is obtained as \cite{art:Dario_Pol}:
\begin{equation}\label{eq:coup_term}
    C_{\mathbf{k}n\nu\sigma j} \simeq -i\left(\frac{\hbar^2e^2}{4m_0\varepsilon_0 }\frac{f_\sigma}{S} \right)^{\frac{1}{2}}
    \sum_{\mathbf{G}} \hat{\mathbf{e}}_\sigma\cdot \mathbf{E}_{\mathbf{k}+\mathbf{G},n}(z_j)F^*_{\mathbf{k}+\mathbf{G},\nu,\sigma,j} \, .
\end{equation}
in which $\mathbf{E}_{\mathbf{k}+\mathbf{G},n}(z_j)$ ($F^*_{\mathbf{k}+\mathbf{G},\nu,\sigma,j}$) is the Fourier component of the electric field (excitonic envelope function) in the $j$-th active layer centered in the $z_j$ position, while $f_\sigma/{S}$ is the oscillator strength per unit surface, in general depending on the polarization. The photon and exciton Fourier components can be retrieved, as previously shown in Secs.~\ref{subsec:GME} and \ref{subsec:sch}. 

The total Hamiltonian, $H=H_{\text{ph}}+H_{\text{ex}}+H_I^{(1)}+H_I^{(2)}$, can be exactly diagonalised by following a generalised Hopfield method \cite{art:Hopfield,art:Dario_Pol}, i.e., by introducing the exciton-polariton destruction bosonic operators $p_{\mathbf{k}}$:
\begin{align}
    \begin{split}
    p_{\mathbf{k}}&= \!\sum_n w_{\mathbf{k}n}\hat{a}_{\mathbf{k}n}
    +\sum_{\nu\sigma j}x_{\mathbf{k}\nu\sigma j }\hat{b}_{\mathbf{k}\nu\sigma j}\\ &+\sum_n y_{\mathbf{k}n}\hat{a}^\dagger_{-\mathbf{k}n}
    + \sum_{\nu\sigma j}z_{\mathbf{k}\nu\sigma j }\hat{b}^\dagger_{-\mathbf{k}\nu\sigma j},
    \end{split}
\end{align}
which diagonalise the new Hamiltonian with the condition $\left[p_{\mathbf{k}},H \right]=\hbar\Omega_{\textbf{k}} p_{\mathbf{k}}$, leading to the construction of a non-Hermitian matrix (the generalized Hopfield operator), as described in detail in Sec.~\ref{sec:appx}. 

As it is well known, the diagonalisation of the Hopfield matrix yields the correct polariton eigenvalues, but not the polariton eigenstates expressed on the exciton/photon basis \cite{Quattropani1986,Ciuti2005}.
However, in the rotating wave approximation (RWA, which is hereby fulfilled), the polaritonic modes can be generally written as superposition of a photonic $\ket{\psi}_{\text{ph}}$ and excitonic $\ket{\phi}_{\text{ex}}$ components: 
\begin{equation}
\ket{\Psi}_{\text{pol}}=\alpha\ket{\psi}_{\text{ph}}+ \beta\ket{\phi}_{\text{ex}}.
\end{equation}
From the last equation, we can easily retrieve the excitonic ($|\beta|^2$) and photonic ($|\alpha|^2$) fractions of the polariton modes, respectively:
\begin{eqnarray}
|\beta|^2=\left|\langle\phi\middle| \Psi\rangle\right|^2,\\
|\alpha|^2=\left|\langle\psi\middle| \Psi\rangle\right|^2,
\end{eqnarray}
with the constraint $|\beta|^2+|\alpha|^2=1$ satisfied under RWA conditions. A direct calculation of these quantities can then be obtained from the Hopfield eigenvectors, as detailed in the Appendix, and can be very useful for a wide range of applications, as it will be shown in the following. For example, polariton-polariton scattering processes, which are responsible for the polariton condensation, strongly depend on the excitonic fraction $|\alpha|^2$, while for long-range transport experiments a lower excitonic fraction is expected to yield lower intrinsic losses \cite{art:Carusotto-Ciuti2013}.

\section{Results }\label{sec:app}

We now report a few examples to show the usefulness of this formulation of radiation-matter interaction in patterned multilayers, which can be applied to a variety of different systems, allowing to directly calculate the complex eigenmodes (i.e., real and imaginary parts) of the hybrid system. We also benchmark the results of this theoretical model through direct comparison with the optical response calculated via rigorous coupled wave analysis (RCWA). For the latter, we use a customized code based on the freely available S$^4$ software \cite{Liu2012} that extends previous RCWA formulations \cite{Whittaker1999,Liscidini2008}.



\subsection{ Vacuum Rabi splitting and number of QWs}

\begin{figure}
    \centering
    \includegraphics[scale=1]{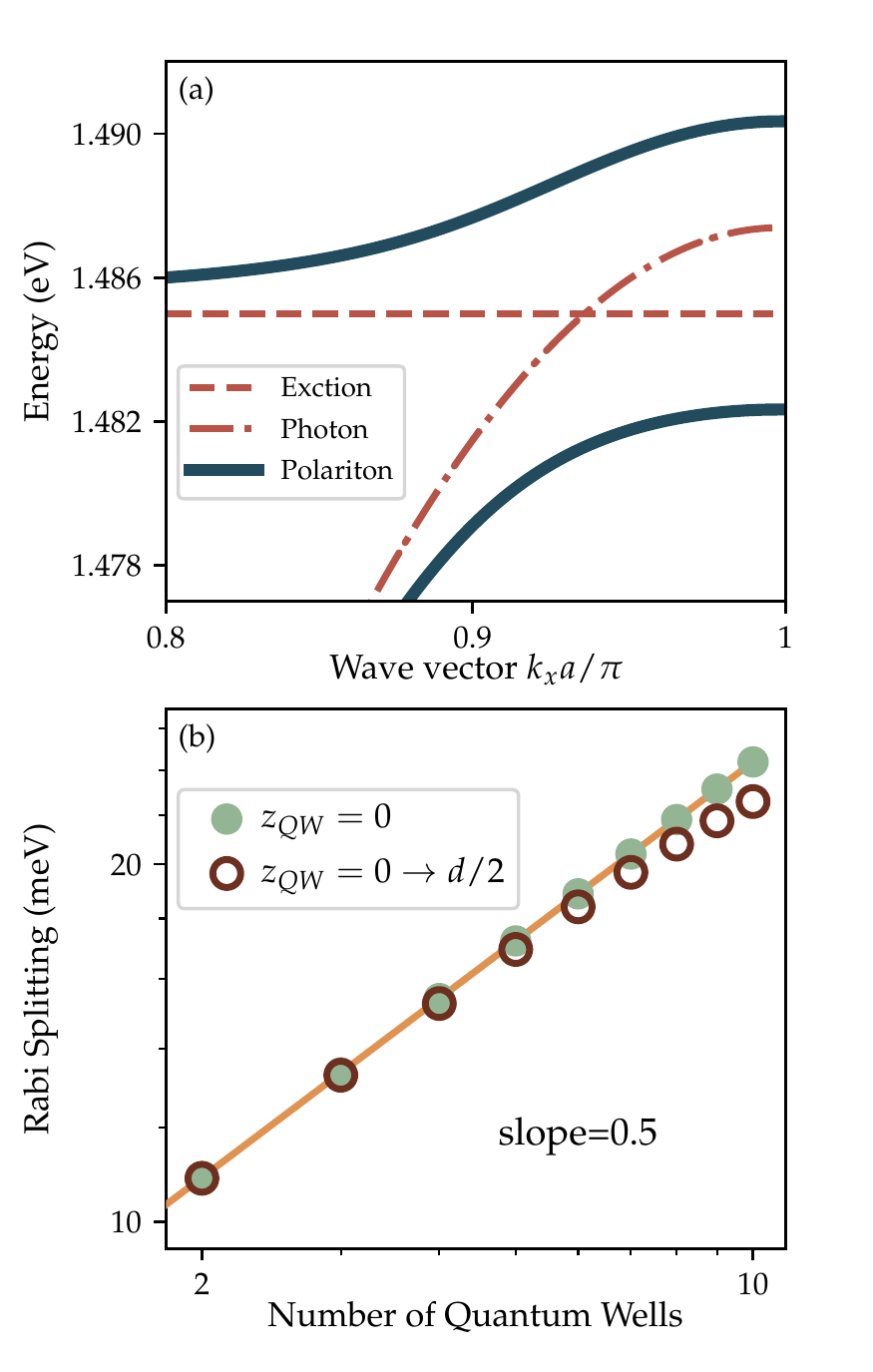}
    \caption{(Colour online) (a) Polaritonic dispersion obtained by the strong coupling between the photonic and excitonic modes for a single quantum well. (b) Log-log plot of the Rabi splitting of the polaritonic dispersion for (a) an increasing number of QWs. Filled circles are relative to QWs placed at the center of the slab (i.e., in $z=0$), empty circles refer to QWs uniformly positioned for increasing $z$ positions, starting from the slab center ($z=0$) to the top ($z=d/2$). The slope of the simulated splittings for all QWs assumed at the centre of the slab is simulated to be $0.500$, as expected.}
    \label{fig:MQWs}
\end{figure}

As a first case study, we analyse a multi-quantum wells layer embedded in a photonic crystal (PhC) slab. A schematic representation of this inorganic semiconductor heterostructure is illustrated in Fig.~\ref{fig:sketch_mqws}. In the following results, the dielectric slab has been assumed to be made of a material with refractive index  $n=3.43$, thickness $d=\SI{658}{\nano\meter} $, and air  holes etched in a square array (see Fig. \ref{fig:sketch_mqws}). We hereby assume a lattice constant  $a=\SI{219}{\nano\meter}$, and holes radius  $r=\SI{75}{\nano\meter}$. The QWs thickness is assumed to be negligible (a good approximation given that III-V semiconductor QWs are usually in the 10-20 nm range thick, i.e. much smaller than their typical exciton transition wavelength), their excitonic resonance is set to \SI{1.485}{eV} with oscillator strength per unit area $f/S=$\SI{4.2e16}{\per\square\meter}. Similar parameters were also considered in \cite{art:Dario_Pol}, but including a single QW. The excitonic modes undergo strong radiation-atter coupling with the truly guided photonic mode, ideally lossless since its dispersion falls below the light line. This result is shown in Fig.~\ref{fig:MQWs}(a), where N=5 photonic and M=5 excitonic modes were kept in the basis to build the Hopfield matrix. In addition, only in-plane polarized excitons are kept in the basis ($\sigma=1,2$), since typical III-V semiconductor QWs possess heavy-hole fundamental excitonic transitions. In the plot, only the polaritonic branches are evidenced, while the exciton center of mass modes have been neglected for easier visualization. We notice that excitonic sub-bands are almost dispersionless and quantized within a \si{\milli\electronvolt} window\cite{art:Dario_Pol}. The Rabi splitting, i.e. the energy separation between the two polariton branches at the resonance condition, can be easily inferred from this plot. 

Here we simulate the same structure for a varying number of QWs, $K\in[1,10]$, in two different configurations. In a first, unrealistic situation, all the QWs are placed at the centre of the slab, i.e. at position $z_{\text{QW}}=0$. In such a case, all QWs interact with the same photonic mode through the same coupling terms \eqref{eq:coup_term}. Hence,  the Rabi splitting is expected to have a $\sqrt{K}$ dependence. The second, realistic configuration is obtained by displacing the QWs in the upper half of the slab, as depicted in Fig. \ref{fig:sketch_mqws}. In this case a deviation from the $\sqrt{K}$ behaviour is expected. This is indeed confirmed by our numerical solutions. In Fig. \ref{fig:MQWs}(b) we plot the Rabi splitting with increasing number of QWs for both configurations, the slope in the log-log plot reveals the predicted $\sqrt{K}$ behavior when all QWs lie at the centre of the slab. Thus, the correct formalization of this problem allows to correctly capture the deviation from the simplistic $\sqrt{K}$ scaling for multi-QW structures, due to the weighted coupling of the guided mode along the slab thickness: QWs that are next to the slab center feel an intense electric field, whose amplitude decreases on moving towards the upper and lower edges such that QWs positioned far from the center are less coupled.

\subsection{Square lattice of perovskite pillars}

\begin{figure}[t]
    \centering
    \includegraphics[width=\linewidth]{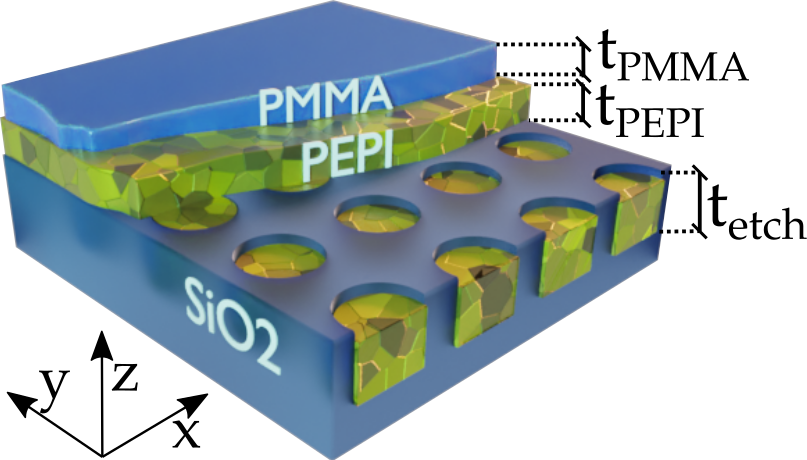}
    \caption{(Colour online) Sketch of perovskite metasurface considered in the main text, and experimentally studied in \cite{art:Ha_My}.} 
    \label{fig:HaiSonPEPI}
\end{figure}

As a second example, we apply this theoretical framework to a 2D layered perovskite, so-called PEPI, supporting a strong excitonic resonance and used as an active material within a periodically textured backbone with square lattice \cite{art:Ha_My}. In this case, the metasurface is composed of a \ch{SiO2} ($n_{\ch{SiO2}}=1.48$) core with etched air holes, in which the PEPI can be infiltrated. Here we consider an etching depth $t_{\text{etch}}=\SI{150}{\nano\meter}$. A residual PEPI layer of thickness $t_{\text{PEPI}}=\SI{30}{\nano\meter}$ lies on top of the etched region, as sketched in Fig~\ref{fig:HaiSonPEPI}. Finally, the device is encapsulated with a layer of a PMMA polymer of thickness  $t_{\text{PMMA}}=\SI{200}{\nano\meter}$ ($n_{\text{PMMA}}=1.49$).
The optical response of this active metasurface can be calculated through RCWA, in which the excitonic response of the PEPI material can be modeled with an isotropic Lortentzian dielectric function:
\begin{equation}\label{eq:Lorentz_PEPI}
    \varepsilon_{\text{PEPI}}\left(E\right)=n^2+\frac{A_\text{ex}}{E_\text{ex}^2-E^2-i\Gamma_\text{ex}E},
\end{equation}
in which $n=2.41$ is the background refractive index, 
$E_\text{ex}=\SI{2.394}{\electronvolt}$ the exciton energy, $\Gamma_\text{ex}=\SI{30}{\milli\electronvolt}$ its linewidth, and $A_\text{ex}=\SI{0.85}{\square\electronvolt}$ is commonly referred to as the ``oscillator strength''. Strictly speaking, however, the oscillator strength in either Lorentz model or quantum theory is a dimensionless quantity. For a layered system with excitonic response, the oscillator strength per unit surface can be related to $A_\text{ex}$ with the following relation \cite{Andreani_Book_2014}:
\begin{equation}
    \frac{f}{S}= \frac{\varepsilon_0m_0 A_\text{ex}}{\hbar^2}\,L \, ,
\end{equation}
where $L$ is the thickness of the active material considered.

\begin{figure}[t]
    \centering
    \includegraphics[scale=1]{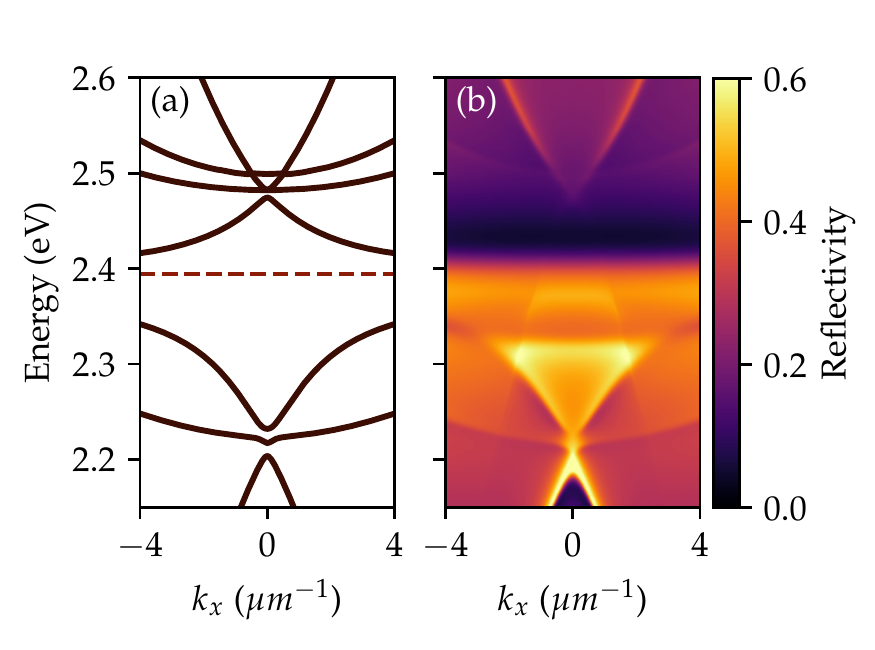}
    \caption{(Colour online) (a) Polaritonic dispersion calculated for the perovskite metasurface with the generalised Hopfield method. The dashed line represents the bare excitonic energy. (b) Reflectivity spectrum calculated for the same structure with the RCWA method.
    }
    \label{fig:disp_HaiSon}
\end{figure}

For the polariton dispersion calculated through diagonalization of the Hopfield matrix, we modeled the active material as a pile up of $M=15$ effective QW layers, each assumed with thickness $L=\left(t_{\text{etch}}+t_{\text{PEPI}}\right)/15=\SI{12}{\nano\meter}$ leading to $f/S=\SI{7.4e18}{\per\square\meter}$ for each effective active layer. Notice that in actual perovskite-based structures the number of QW layers is much larger, which we could model exactly, in principle; however, we realize that no appreciable change would be obtained. Depending on the $z$-position of the effective active layer, excitons will experience a different 2D potential profile in the $xy$-plane, which affects the excitonic quantized energies, $E_{\mathbf{k}}$, due to different potential terms $V\left(\mathbf{G}-\mathbf{G}'\right)$ in Eq.~\eqref{eq:sch_rec}. In addition, we assume $\gamma_{\text{ex}}=\Gamma_{\text{ex}}/2=15$ meV as the exciton imaginary part in the Hopfield matrix. For the photonic calculations only odd modes with respect to the plane of incidence were taken into account, namely, transverse electric (TE, i.e., s-polarised) modes. The resulting polaritonic dispersion is shown in Fig.~\ref{fig:disp_HaiSon}(a). In particular, the extrapolated excitonic fraction at the bottom of the uppermost band is $0.35$, as it was previously reported \cite{art:Ha_My}, which confirms the accuracy of the method. 

For direct comparison and benchmarking of our model, we also report the optical reflectivity spectra calculated through RCWA in Fig.~\ref{fig:disp_HaiSon}(b). Guided resonances appear as narrow Fano-like features on a reflection background, which can be clearly evidenced in the color scale plot. The angular dispersion of these features (which is reported as an energy-wave vector dependence) can be put in one-to-one correspondence with the calculated polariton modes in Fig.~\ref{fig:disp_HaiSon}(a). In order to compare the two simulations, we slightly shifted the photonic bands calculated by GME method and then used in the Hopfield matrix, by assuming a PEPI material index $n=2.46$, as  compared to the $n=2.41$ of the Lorentzian response in Eq. (\ref{eq:Lorentz_PEPI}) that we employed in the RCWA simulations. Apart from the approximation of a layered sequence of 15 QWs used to mimic the PEPI (a uniform and isotropic layer is used, instead, in the RCWA simulation), no other adjustable parameter was employed and the two simulations are perfectly consistent with each other. We highlight the overall agreement over a large spectral window, which justifies a posteriori our assumption of using 15 layers, and it shows the great potential power of the Hopfield method. In fact, numerical diagonalization of the Hopfield matrix (with $N=8$ and $M=10$) only takes a few minutes to recover the results in Fig.~\ref{fig:disp_HaiSon}(a) on a normal Desktop processor. Correspondingly, RCWA simulations are known to be computationally more demanding, and typically require about one order of magnitude longer simulation times on comparable hardware and for comparable parameters (number of plane waves in the Fourier basis).


\begin{figure}[t]
    \centering
    \includegraphics[width=0.5\textwidth]{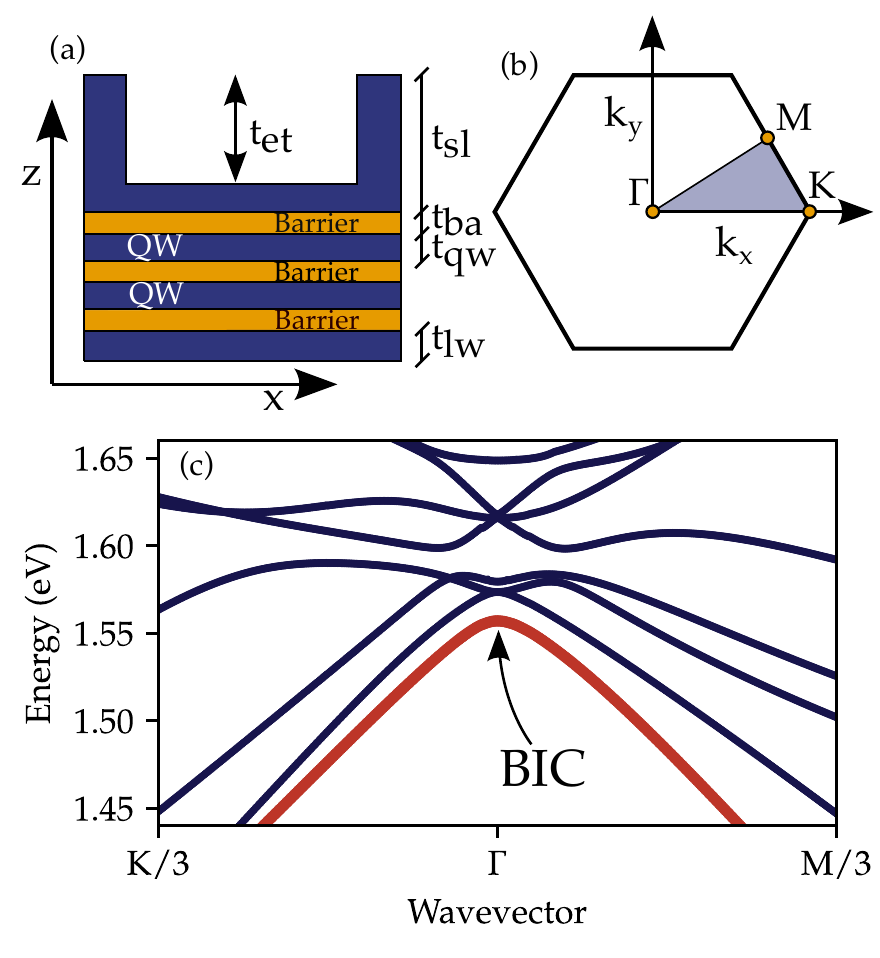}
    \caption{(Colour online) (a) Cross-sectional view ($xz$ plane) of the partially etched dielectric slab with two quantum well layers in the unpatterned region. (b) Brillouin zone for a triangular lattice, in which high-symmetry points have been highlighted. (c) Purely photonic band dispersion around the $\Gamma$ point of the partially etched structure with triangular lattice of holes, where M$/3$ (K$/3$) represents one third of the $\Gamma$$\to$M ($\Gamma$$\to$K) high-symmetry line. A bound-state in the continuum is found at the $\Gamma$ point of the first band. }
    \label{fig:tri_phot}
\end{figure}

\subsection{Coupling to bound-states in the continuum}

\begin{figure}[t]
    \centering
\includegraphics[width=0.5\textwidth]{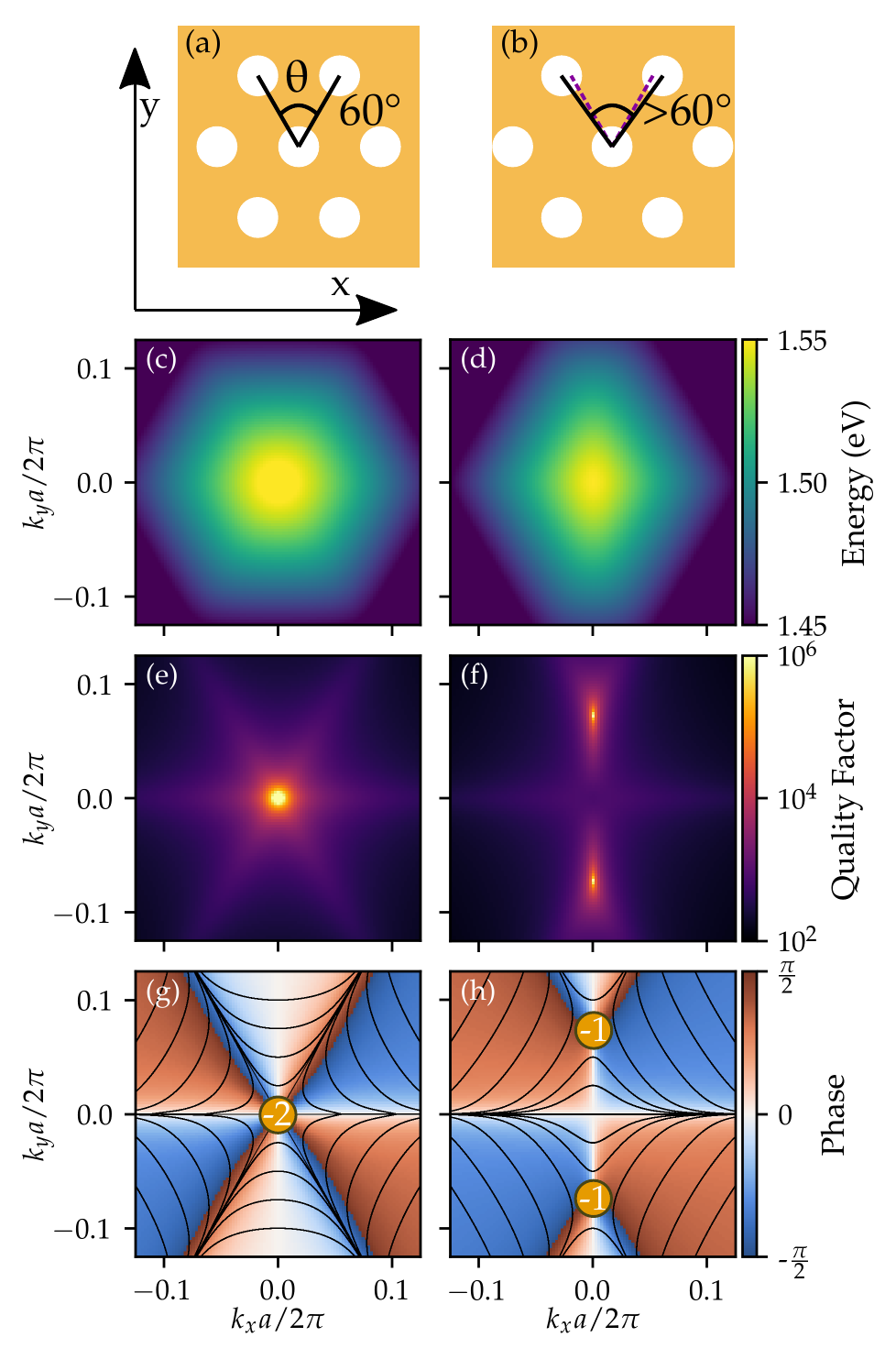}
    \caption{(Colour online) Left panels refer to the first photonic band of the standard triangular lattice shown in (a), right panels refer to the same band for the lattice with broken symmetry shown in (b). (c)-(d) Energy dispersion around $\Gamma$ represented as color scale plot as a function of ($k_x,k_y$). (e)-(f) Quality factor calculated for the bands shown in (c)-(d); the BIC of the triangular lattice in (e) splits into two BICs in (f), since the triangular symmetry is broken. (g)-(h) Polarisation angle of electric field and associated topological charges. Black lines represent the polarisation vector.}
    \label{fig:phot_2D}
\end{figure}

Bound-states in the continuum (BIC)s are peculiar solutions of a wave equation that remain indefinitely confined in a resonant system despite lying in a continuum spectrum, where modes are supposed to radiate. This paradigm has been recently highlighted in Photonics as well \cite{Hsu2016,Azzam2021}, in which non-trivial topological properties were evidenced\cite{Zhen2014,Zhang2018}. Polariton BICs have recently received significant attention in different physical realizations\cite{art:Koshelev18,Kravtsov2020,art:Ha_My,art:Koreani21,art:Ardizzone2022}, also in connection to their topological properties. Recently, it has been demonstrated that photonic BICs can be tailored by exploiting symmetry breaking of the underlying structure \cite{art:Notomi}. Here we used this degree of freedom to flexibly shift the BIC condition along high symmetry lines in reciprocal space, thus enabling the strong coupling between an off-$\Gamma$ photonic BIC and the excitonic resonance. To the best of our knowledge, this is the first application of such concepts to polariton physics.

\begin{figure}[t]
    \centering
    \includegraphics[width=0.5\textwidth]{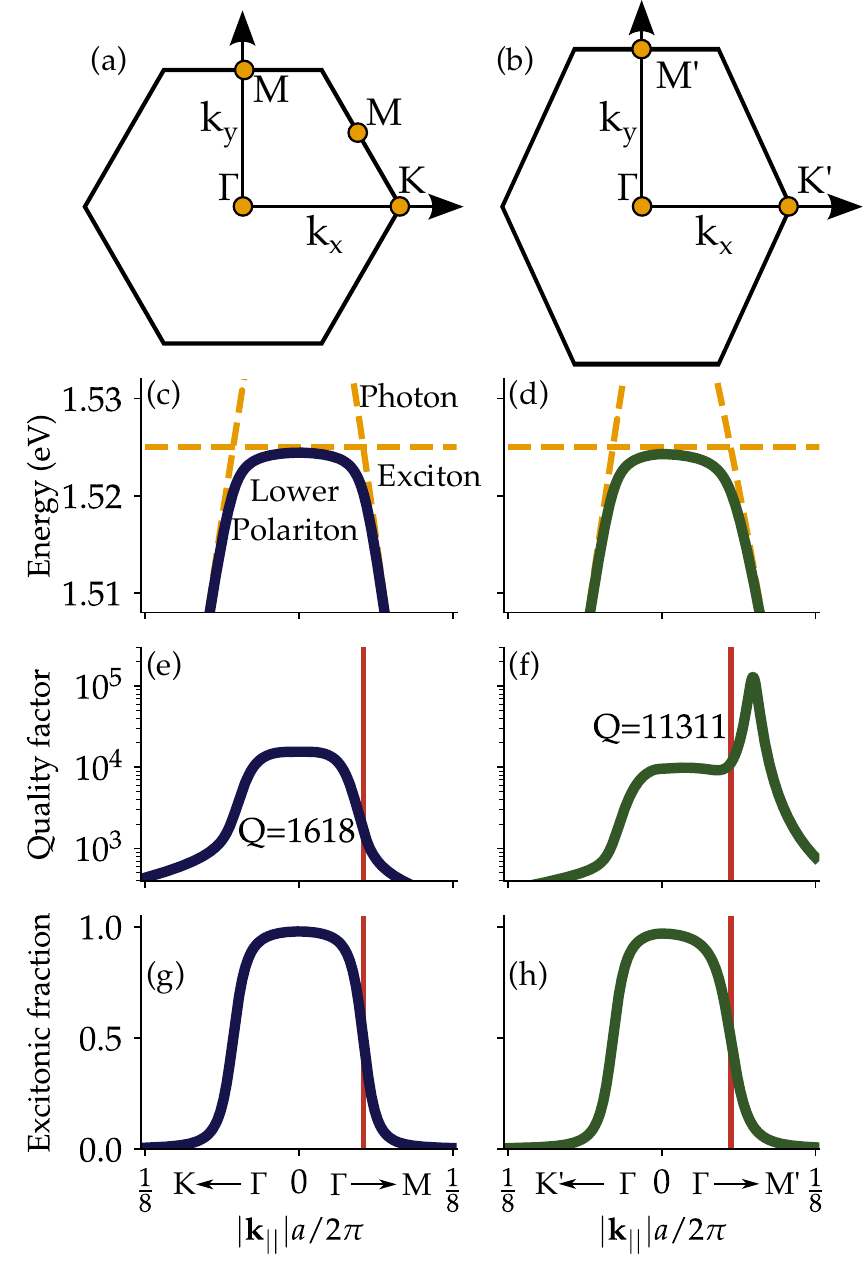}
    \caption{(Colour online) Lower polariton mode calculated along two symmetry lines. Left panels refer to the standard triangular lattice of Fig.~\ref{fig:phot_2D}(a) whose Brillouin zone is shown in (a), right panels refer to the lattice with $C_6$ symmetry broken (while the $C_2$ symmetry is preserved), i.e. corresponding to the photonic bands of Fig.~\ref{fig:phot_2D}(b), whose Brillouin zone is shown in (b). The vertical lines in (e)-(h) denote the wave vector for which the excitonic fraction is $|\alpha|^2=0.5$.
    (c)-(d) Lower polariton branches obtained from excitonic and photonic dispersions along the symmetry lines indicated in (a)-(b). (e)-(f) Q factor of the polaritonic dispersions shown in (c)-(d); the Q factors at excitonic fraction $|\alpha|^2=0.5$ are explicitly reported in the figures. (g)-(h) Excitonic fraction corresponding to the lower polariton bands in (c)-(d).
    }
    \label{fig:pol_60_62}
\end{figure}

The starting structure is a dielectric slab partially etched with two active QW layers embedded, as pictured in Fig.~\ref{fig:tri_phot}(a). We considered air holes etched in a triangular lattice. The main parameters are the thicknesses of the upper slab, QWs, barrier and lower film, respectively, which we fix as: $t_{\text{sl}}=\SI{132.5}{\nano\meter}$, $t_{\text{qw}}=\SI{20}{\nano\meter}$, $t_{\text{ba}}=\SI{10}{\nano\meter}$, $t_{\text{lw}}=\SI{22.5}{\nano\meter}$. The etching depth is assumed $t_{\text{et}}=\SI{110}{\nano\meter}$. The lattice constant is set to $a=\SI{305}{\nano\meter}$ and the hole radius to $r=\SI{106}{\nano\meter}$. The refractive index of the barriers is $n_{\text{ba}}=3.3$, while for all the other layers the assumed refractive index is $n_{\text{qw}}=3.56$. We notice that these are typical values for material platforms based on \ch{Al_{0.4}Ga_{0.6}As} barriers and \ch{GaAs} QWs in the considered energy range. The exciton energy for such QWs is typically centered at $E=\SI{1.525}{\electronvolt}$. 

\begin{figure}[t]
    \centering
    \includegraphics[width=0.5\textwidth]{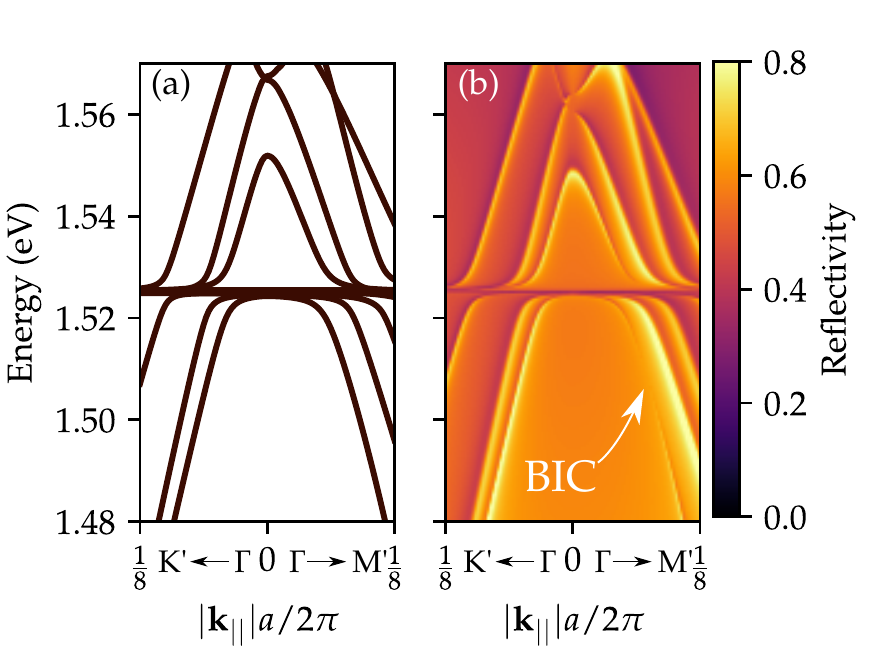}
    \caption{(Colour online) Real part of polariton eigenmodes (a) calculated by diagonalization of the Hopfield matrix for the lattice of Fig.~\ref{fig:pol_60_62}(b), and corresponding RCWA simulation (b) of angle-resolved reflectivity spectra, clearly showing the off-$\Gamma$ BIC.
    }
    \label{fig:GaAs_S4_Legume}
\end{figure}

The purely photonic dispersion calculated by GME along high-symmetry lines around the $\Gamma$ point is shown in Fig.~\ref{fig:tri_phot}(c). For the present example, we considered the first band at lowest energy, displaying a BIC in $\Gamma$. The latter is clearly identified in the GME calculation, since its imaginary part calculated by perturbation theory through Eq.~\ref{eq:im-part} is identically zero. This band is a good candidate since the at-$\Gamma$ BIC has a topological charge $q=-2$, as defined with an integral on a closed loop $C$ in the $\mathbf{k}_{||}=\left(k_x,k_y\right)$ space around the BIC\cite{art:Notomi}:
\begin{equation}
    q=\frac{1}{2\pi}\int_C \nabla_{\mathbf{k}_{||}}\phi\left(\mathbf{k}_{||}\right)\cdot d\mathbf{k}_{||} \, ,
\end{equation}
where $\phi\left(\mathbf{k}_{||}\right)$ is the polarisation angle of the electric field in the slab. This BIC can be split into two off-$\Gamma$ BICs, each with a topological charge $q=-1$, by just breaking the $C_6$ symmetry of the triangular lattice, i.e., its invariance under a $\pi/3$ rotation \cite{art:Notomi}. This opens up the possibility to achieve strong radiation-matter coupling between the excitonic resonance and a photonic state with vanishing losses at finite wave vector. As a result, a polaritonic mode with extremely high quality (Q) factor, and consequently extremely low losses (i.e., long lifetime), will form at the same wave vector. Notice that the Q factor is calculated for each GME mode as $Q_{\mathbf{k}n}=\omega_{\mathbf{k}n}/2 \gamma_{\mathbf{k}n}$; hence, the Q factor diverges at the exact BIC condition, but in the color scale plot this is cut to an upper bound of $10^6$ to be displayed in log-scale.

To break the $C_6$ symmetry it is sufficient to slightly shift the holes within the elementary unit cell, e.g., by modifying the angle $\theta$ defined in Fig.~\ref{fig:phot_2D}(a). Here, the holes have been shifted in such a way to form an angle $\theta=\SI{62.45}{\degree}$, as shown schematically in Fig.~\ref{fig:phot_2D}(b). Under such conditions, the  at-$\Gamma$ BIC with energy $E\approx\SI{1.56}{\electronvolt}$, characterised by a diverging Q factor, splits into two BICs at $k_ya/2\pi\approx\pm0.073$ and energy $E\approx\SI{1.52}{\electronvolt}$, as shown in Figs. \ref{fig:phot_2D}(e)-(f). 

\begin{figure}[t]
    \centering
    \includegraphics[width=0.45\textwidth]{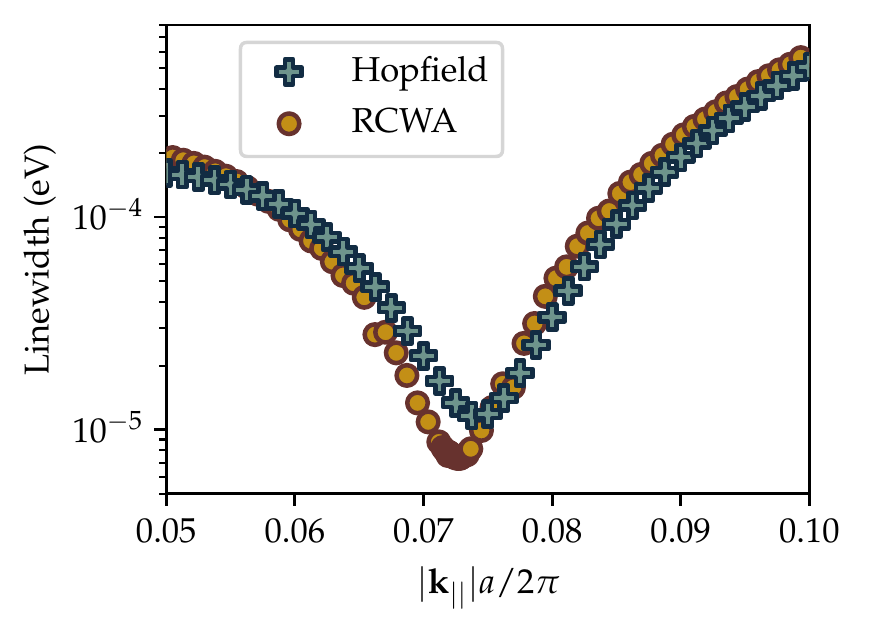}
    \caption{(Colour online) Linewidth associated to the lowest energy polaritonic band (where the polariton BIC is highlighted in Fig.~\ref{fig:GaAs_S4_Legume}b). In From the RCWA simulated absorption spectra, the linewidth is extracted from Lorentzian fits of the resonant features; in the generalised Hofpield method the linewidth is explicitly obtained after diagonalization by taking twice the imaginary part of the corresponding complex energy eigenvalue.
    }
    \label{fig:pol_loss}
\end{figure}

These two off-$\Gamma$ BICs can undergo strong coupling with the exciton resonance at $E=\SI{1.525}{\electronvolt}$, thus forming off-$\Gamma$ polariton BICs. This is explicitly shown shown in Fig.~\ref{fig:pol_60_62}, where we show the results of the Hopfield matrix diagonalization for the vertical heterostructure of Fig.~\ref{fig:tri_phot}(a) and a triangular lattice with $a=302$ nm. The left panels refer to the lattice with $\theta=\SI{60}{\degree}$, while the right ones correspond to $\theta=\SI{62.45}{\degree}$. 

Figures~\ref{fig:pol_60_62}(c)-(d) show the lower polariton branches around $\Gamma$ and along the symmetry directions indicated in panels (a)-(b), derived from the strong coupling of the photonic and excitonic resonances for both configurations. Despite the two polaritonic dispersions look very similar, they strongly differ in their loss properties. In fact, the polariton branch at excitonic fraction $|\alpha|^2=0.5$ exhibits a Q factor that is almost seven times greater when the exciton is coupled to the BIC, as evidenced in Figs.~\ref{fig:pol_60_62}(e)-(f). This implies a polariton lifetime that is almost seven times longer for the polariton-BIC as compared to the lossy polariton mode at the same wave vector. The off-$\Gamma$ polariton BIC is even more clearly identified by a strongly peaked Q-factor in panel (f), whose finite value is bound by excitonic losses in the corresponding Hopfield eigenvalue even if the purely photonic Q-factor diverges. The corresponding excitonic fraction is plotted in Figs.~\ref{fig:pol_60_62}(g)-(h). These results are particularly relevant, since they shown that photonic crystal polariton losses can be fully engineered within the Brillouin zone through slight distortions of the lattice. \\
Finally, the topological charge conservation is demonstrated from the polarisation maps reported in Figs.~\ref{fig:phot_2D}(g)-(h). While these results have been obtained for the purely photonic modes, their topological properties can be transferred one-to-one to the polariton BICs, as already shown \cite{art:Ha_My}.

As a final result, we show the direct comparison of Hopfield diagonalization and RCWA simulation of the optical response of such a multilayered structure. This is reported in Fig.~\ref{fig:GaAs_S4_Legume}, where the polariton dispersion (i.e., real part of the complex energy eigenvalues) obtained from the diagonalization of the Hopfield matrix in panel (a) are shown to be in one-to-one correspondence with the reflectivity features in the RCWA simulated spectra in panel (b), for the very same structure parameters of Fig.~\ref{fig:pol_60_62} corresponding to the distorted triangular lattice. The off-$\Gamma$ polariton BIC is clearly seen in  Fig.~\ref{fig:GaAs_S4_Legume}(b) from the vanishing reflectivity signal on top of the background. The overall agreement is remarkable, highlighting once more the effectiveness of the generalized Hopfield method. We just notice that a slightly different refractive index for the QW material was assumed in the GME calculation of purely photonic modes ($n_{\text{qw}}=3.56$), as compared to the RCWA simulation ($n_{\text{qw}}=3.53$), to slightly adjust a few meV  blue shift of the whole spectrum that is typically obtained in GME as compared to RCWA. \\
The comparison between the two approaches is further benchmarked by explicitly showing the polariton mode linewidth around the BIC condition, in Fig.~\ref{fig:pol_loss}. Despite a small discrepancy due to the BIC condition occurring at a slightly different point in reciprocal space, the overall agreement is remarkable. We also notice that the Lorentzian fitting of the RCWA absorption spectra (not shown) allows to evidence a clear saturation of losses et the off-$\Gamma$ BIC. While the fitting is not possible at the exact BIC condition (due to the vanishing resonant response, which is due to the radiative losses becoming identically zero), the residual loss can be evidently attributed to the non-radiative component of excitonic losses. In the diagonalized Hopfield matrix, the latter is automatically contained into the imaginary part of the corresponding eigenvalue, which is due to the excitonic fraction of the lower polariton BIC mode even when the photonic losses become identically zero. This figure clarifies the connection between the complex mode eigenvalues obtained within a coupled oscillators model, and the optical spectra derived from either RCWA or experimental data.   

\section{Conclusions}

We have reported a quantum theory of photonic crystal polaritons, based on a generalization of the Hopfield matrix that takes into account the in-plane periodic modulation of a multilayered structure containing active layers with 2D excitonic response. We have shown that this approach is extremely general and flexible, allowing to simultaneously capture the essential physics of very diverse material platforms and heterostructures, from inorganic multi-quantum wells with partial etching, to infiltrated metasurfaces used as a backbone for active media, and deposited thin layers, e.g., two-dimensional materials, on top of patterned gratings. We have shown a few examples of application of the method, all connected to the polariton physics and strong radiation-matter coupling between excitonic and photonic modes, such as the correct derivation of the Rabi splitting scaling as a function of the number of active layers distributed along the photonic structure thickness, the dispersion of perovskite based polaritons, and the non-trivial topological properties of polariton bound states in the continuum. Interestingly, our results can be put in one-to-one correspondence with quantitative information derived from the optical spectra. We have shown that rigorous coupled wave simulations of optical reflectivity (or absorption) match remarkably well the complex eigenvalues obtained from diagonalizing the Hopfield matrix, by comparing either the real part to the energy of the resonant features, or the imaginary part to half their linewidth. This numerically fast and efficient approach promises then to serve as a useful tool to analyse the physics of polariton eigenmodes, to be used as an alternative or in combination with either numerical results obtained with classical approaches, or experimental spectra obtained by optical excitation of such heterostructures. 
It can be the basis for describing nonlinear polariton interactions in photonic crystal structures.
Furthermore, the very same concepts could be extended to deal with strongly coupled electromagnetic excitations in other contexts, from phonon to intersubband polaritons.

\acknowledgments
This work was partly supported by the Italian Ministry of Research (MUR) through PRIN project ``INPhoPOL''.
H.S.N is supported by the French National Research Agency (ANR) under the project POPEYE (ANR-17-CE24-0020) and the IDEXLYON from Universit\'e de Lyon, Scientific Breakthrough project TORE within the Programme Investissements d’Avenir (ANR-19-IDEX-0005).
Very useful discussions with Davide Nigro and Luca Zagaglia are gratefully acknowledged. Some of the colormaps used in the manuscript are designed by {B}. {R}. {M}illis, and they can be found at the link: https://github.com/BlakeRMills/MetBrewer.

\begin{widetext}
\section{Appendix}\label{sec:appx}

The interaction Hamiltonian can be written in terms of photonic $\hat{a}$ and excitonic $\hat{b}$ operators as:
\begin{align}
\begin{split}
    H_{\mathbf{k}}= \underbrace{\sum_n\hbar\omega_{\mathbf{k}n}
    \hat{a}_{\mathbf{k}n}^\dagger\hat{a}_{\mathbf{k}n}}_{H_{\text{ph}}}  +
    \underbrace{\sum_{\nu,\sigma,j}E_{\mathbf{k}\nu\sigma j }
    \hat{b}_{\mathbf{k}\nu\sigma j}^\dagger\hat{b}_{\mathbf{k}\nu\sigma j}}_{H_{\text{ex}}}
    +\underbrace{i\sum_{n,\nu,\sigma,j}C_{\mathbf{k}n\nu\sigma j}
    (\hat{a}_{\mathbf{k}n}+\hat{a}_{-\mathbf{k}n}^\dagger)(\hat{b}_{-\mathbf{k}\nu\sigma j}-\hat{b}_{\mathbf{k}\nu\sigma j}^\dagger)}_{H_\text{I}^{(1)}} \\
    +\underbrace{\sum_{\mathbf{k},nn',\nu,\sigma,j }D_{\mathbf{k}nn'\nu\sigma j }
    (\hat{a}_{-\mathbf{k}n}+\hat{a}_{\mathbf{k}n}^\dagger)(\hat{a}_{\mathbf{k}n'}+\hat{a}_{-\mathbf{k}n'}^\dagger)}_{H_\text{I}^{(2)}}.
\end{split}
\end{align}
where $n\in\left[1,N\right]$ is the photonic band index , $\nu\in\left[1,M\right]$ is the excitonic band index and $j\in\left[1,K\right]$ is the index of the active layer.
Introducing the polaritonic operators defined as:
\begin{equation}
    p_{\mathbf{k}}= \sum_n w_{\mathbf{k}n}\hat{a}_{\mathbf{k}n}
    +\sum_{\nu\sigma j}x_{\mathbf{k}\nu\sigma j }\hat{b}_{\mathbf{k}\nu\sigma j}
    +\sum_n y_{\mathbf{k}n}\hat{a}^\dagger_{-\mathbf{k}n}
    + \sum_{\nu\sigma j}z_{\mathbf{k}\nu\sigma j }\hat{b}^\dagger_{-\mathbf{k}\nu\sigma j}.
\end{equation}
The Hamiltonian can be diagonalised in terms of the polaritonic operators $p_{\mathbf{k}}$ by imposing $\left[p_{\mathbf{k}},H \right]=\hbar\Omega_{\textbf{k}} p_{\mathbf{k}}$ where $\hbar\Omega_{\textbf{k}}$ are the polaritonic energies.
It is convenient to evaluate the commutator with each Hamiltonian block separately:
\begin{equation}
    \left[p_{\mathbf{k}},H_{\text{ph}} \right]=
    \sum_n\hbar\omega_{\mathbf{k}n}(w_{\mathbf{k}n} \hat{a}_{\mathbf{k}n}-y_{\mathbf{k}n} \hat{a}^\dagger_{-\mathbf{k}n} ),
\quad
    \left[p_{\mathbf{k}},H_{\text{ex}} \right]=
    \sum_{\nu,\sigma,j} E_{\mathbf{k}\nu\sigma j}(x_{\mathbf{k}\nu\sigma j} \hat{b}_{\mathbf{k}\nu\sigma j}-z_{\mathbf{k}\nu\sigma j} \hat{b}^\dagger_{-\mathbf{k}\nu\sigma j} ),
\end{equation}
\begin{align}
\begin{split}
    \left[p_{\mathbf{k}},H_{I}^{(1)} \right]=
    i\sum_{n,\nu,\sigma,j}C_{-\mathbf{k}n\nu\sigma j}
    \left\{(w_{\mathbf{k}n}-y_{\mathbf{k}n})
    (\hat{b}_{\mathbf{k}\nu\sigma j}-\hat{b}^\dagger_{-\mathbf{k}\nu\sigma j})
    \right\}
    -i \sum_{n,\nu,\sigma,j}C_{\mathbf{k}n\nu\sigma j}
    \left\{(x_{\mathbf{k}\nu\sigma j}+z_{\mathbf{k}\nu\sigma j})
    (\hat{a}_{\mathbf{k}n}+\hat{a}^\dagger_{-\mathbf{k}n})
,    \right\}
\end{split}
\end{align}
\begin{align}
\begin{split}
    \left[p_{\mathbf{k}},H_{I}^{(2)} \right]=
    \sum_{\nu,\sigma,j,n,n'}\Bigl\{(w_{\mathbf{k}n}-y_{\mathbf{k}n} )\bigl(
    D_{-\mathbf{k} nn'\nu\sigma j}\hat{a}_{\mathbf{k}n'}
    +
    D_{\mathbf{k} nn'\nu\sigma j}\hat{a}_{\mathbf{k}n'}
    +
    D_{\mathbf{k} nn'\nu\sigma j}\hat{a}^\dagger_{-\mathbf{k}n'}
    +
    D_{-\mathbf{k}nn'\nu\sigma j}\hat{a}^\dagger_{-\mathbf{k}n'}\bigr)\Bigr\}.
\end{split}
\end{align}
In the absence of losses, the Hamiltonian is Hermitian, which  implies $C_{\mathbf{k}}=C_{-\mathbf{k}}^* $ and $D_{\mathbf{k}}=D_{-\mathbf{k}}^*$, moreover $D_{\mathbf{k}mn \nu\sigma j}=D_{\mathbf{k}nm\nu\sigma j}$.  
Now, in order to simplify the linear problem $\left[p_{\mathbf{k}},H \right]=\hbar\Omega_{\textbf{k}} p_{\mathbf{k}}$ we calculate the following commutators:
\begin{align}
    \left[\hat{a}^\dagger_{\mathbf{k}n} ,\left[p_{\mathbf{k}},H \right]\right]=\hbar\Omega_{\textbf{k}}
    \left[\hat{a}^\dagger_{\mathbf{k}n} , p_{\mathbf{k}}\right],\\
     \left[\hat{a}_{\mathbf{k}n} ,\left[p_{\mathbf{k}},H \right]\right]=\hbar\Omega_{\textbf{k}}
    \left[\hat{a}_{\mathbf{k}n} , p_{\mathbf{k}}\right],\\
    \left[\hat{b}^\dagger_{\mathbf{k}\nu\sigma j} ,\left[p_{\mathbf{k}},H \right]\right]=\hbar\Omega_{\textbf{k}}
    \left[\hat{b}^\dagger_{\mathbf{k}\nu\sigma j} , p_{\mathbf{k}}\right],\\
     \left[\hat{b}_{\mathbf{k}\nu\sigma j} ,\left[p_{\mathbf{k}},H \right]\right]=\hbar\Omega_{\textbf{k}}
    \left[\hat{b}_{\mathbf{k}\nu\sigma j} , p_{\mathbf{k}}\right],
\end{align}
which lead to the following set of equations where $\Re$ denotes the real part:
\begin{align}
\begin{split}
    w_{\mathbf{k}n}\hbar\omega_{\mathbf{k}}-i\sum_{\nu,\sigma,j}C_{\mathbf{k}\nu\sigma nj}x_{\mathbf{k}\nu\sigma j}-i\sum_{\nu,\sigma,j}C_{\mathbf{k}\nu\sigma nj}z_{\mathbf{k}\nu\sigma j}
    +\sum_{n',\nu,\sigma,j}2\Re\left[ D_{\mathbf{k}nn'\nu\sigma  j}\right]w_{\mathbf{k}n'}-
    \sum_{n',\nu,\sigma,j}2\Re\left[ D_{\mathbf{k}nn'\nu\sigma  j}\right]y_{\mathbf{k}n'}=
    \hbar\Omega_{\mathbf{k}}w_{\mathbf{k}n},
\end{split}
\end{align}
\begin{align}
\begin{split}
    x_{\mathbf{k}\nu\sigma j}E_{\mathbf{k}\nu\sigma j}+
    i\sum_nC^*_{\mathbf{k}n\nu\sigma j}w_{\mathbf{k}n\nu\sigma j}-i\sum_nC^*_{\mathbf{k}n\nu\sigma j}y_{\mathbf{k}n\nu\sigma j}
    =\hbar\Omega_{\mathbf{k}}x_{\mathbf{k}\nu\sigma j},
\end{split}
\end{align}
\begin{align}
\begin{split}
    -y_{\mathbf{k}n}\hbar\omega_{\mathbf{k}}-i\sum_{\nu,\sigma,j}C_{\mathbf{k}\nu\sigma nj}x_{\mathbf{k}\nu\sigma j}-i\sum_{\nu,\sigma,j}C_{\mathbf{k}\nu\sigma nj}z_{\mathbf{k}\nu\sigma j}
    +\sum_{n',\nu,\sigma,j}2\Re\left[ D_{\mathbf{k}nn'\nu\sigma  j}\right]w_{\mathbf{k}n'}-
    \sum_{n',\nu,\sigma,j}2\Re\left[ D_{\mathbf{k}nn'\nu\sigma j}\right]y_{\mathbf{k}n'}=
    \hbar\Omega_{\mathbf{k}}y_{\mathbf{k}n},
\end{split}
\end{align}
\begin{align}
\begin{split}
    -z_{\mathbf{k}\nu\sigma j}E_{\mathbf{k}\nu\sigma j}
    -i\sum_nC^*_{\mathbf{k}n\nu\sigma j}w_{\mathbf{k}n\nu\sigma j}+i\sum_nC^*_{\mathbf{k}n\nu\sigma j}y_{\mathbf{k}n\nu\sigma j}
    =\hbar\Omega_{\mathbf{k}}z_{\mathbf{k}\nu\sigma j}.
\end{split}
\end{align}
This set of equations can be written as a linear eigenvalue problem with a generalised Hopfield matrix:
\begin{equation}\label{mat:hop}
M=
\begin{pmatrix}
\boldsymbol{\omega}+2\Re\mathbf{D} & -i\mathbf{C}_1 &-i\mathbf{C}_2 & \dots & -i\mathbf{C}_K&-2\Re\mathbf{D} & -i\mathbf{C}_1 &-i\mathbf{C}_2 & \dots & -i\mathbf{C}_K \\
i\mathbf{C}_1^\dagger & \mathbf{E}_1 & \mathbf{0}&\dots & \mathbf{0}&-i\mathbf{C}_1^\dagger & \mathbf{0}& \mathbf{0}&\dots & \mathbf{0} \\
i\mathbf{C}_2^\dagger &  \mathbf{0} & \mathbf{E}_2&\dots & \mathbf{0}&-i\mathbf{C}_2^\dagger & \mathbf{\mathbf{0}} & \mathbf{0}&\dots & \mathbf{0} \\
\vdots & \vdots  &\vdots &\ddots & \mathbf{0}&\vdots &\vdots &\vdots&\ddots & \mathbf{0} \\
i\mathbf{C}_K^\dagger & \mathbf{0} & \mathbf{0} &\dots&\mathbf{E}_K&-i\mathbf{C}_K^\dagger & \mathbf{0} & \mathbf{0} &\dots&\mathbf{0} \\ 
2\Re\mathbf{D} & -i\mathbf{C}_1 &-i\mathbf{C}_2 & \dots & -i\mathbf{C}_K&-\boldsymbol{\omega}-2\Re\mathbf{D} & -i\mathbf{C}_1 &-i\mathbf{C}_2 & \dots & -i\mathbf{C}_K \\
-i\mathbf{C}_1^\dagger &\mathbf{0} & \mathbf{0}&\dots & \mathbf{0}&+i\mathbf{C}_1^\dagger &  -\mathbf{E}_1& \mathbf{0}&\dots & \mathbf{0} \\
-i\mathbf{C}_2^\dagger & \mathbf{0} & \mathbf{0}&\dots & \mathbf{0}&+i\mathbf{C}_2^\dagger & \mathbf{0} &  -\mathbf{E}_2&\dots & \mathbf{0} \\
\vdots & \vdots &\vdots&\ddots & \mathbf{0}&\vdots & \vdots & \vdots&\ddots & \mathbf{0} \\
-i\mathbf{C}_K^\dagger & \mathbf{0} & \mathbf{0} &\dots&\mathbf{0}&+i\mathbf{C}_K^\dagger & \mathbf{0} & \mathbf{0} &\dots&-\mathbf{E}_K \\
\end{pmatrix},
\end{equation}

\end{widetext}
whose eigenvalues correspond to the polaritonic energies.
Note that that j-th quantum well is coupled to the photonic modes by the $\mathbf{C}_j$ block.

Losses can be included a posteriori by assuming complex photon and exciton energies. Specifically, the blocks composing the generalised Hopfield matrix \eqref{mat:hop} including losses are: the $N\times N$ matrix (photon losses included as imaginary parts of the eigenvalues, as discussed in Sec.~\ref{subsec:GME})
\begin{equation}
    \boldsymbol{\omega}+2\Re\mathbf{D}=[\![ \hbar\tilde{\omega}_n \delta_{n,n'}+\sum_{\nu,\sigma,j}2\Re\left[  D_{nn' \nu\sigma j}\right]  ]\!],
\end{equation}
where $\tilde{\omega}_n= \omega_n - i\gamma_n$,
the $3M\times 3M$ matrices (exciton losses included as imaginary parts of the exciton energies)
\begin{equation}
    \mathbf{E}_j=[\![\tilde{E}_{\nu\sigma}^j\delta_{(\nu\sigma),(\nu'\sigma')} ]\!],
\end{equation}
where $\tilde{E}_{\nu\sigma}^j=E_{\nu\sigma}^j - i\gamma_{\mathrm{ex}}$, 
the $N\times 3M$ matrices 
\begin{equation}
    \mathbf{C}_j=[\![ C_{n,\nu,\sigma}^j ]\!],
\end{equation}
and the zero-filled $3M\times 3M$ matrix denoted as $\mathbf{0}$. The notation of the Hopfield matrix \eqref{mat:hop} can be further simplified by introducing $\mathbf{\Gamma}$ and $\mathbf{\Xi}$ defined as:
\begin{equation}
\mathbf{\Gamma} = 
\begin{bmatrix}
\mathbf{C}_1 & \mathbf{C}_2 &\dots &\mathbf{C}_K\\
\end{bmatrix},\quad
\mathbf{\Xi} = 
\begin{bmatrix}
\mathbf{E}_1 & \mathbf{0} &\dots &\mathbf{0}\\
\mathbf{0}& \mathbf{E}_2 &\dots &\mathbf{0}\\
\vdots& \vdots &\ddots &\mathbf{0}\\
\mathbf{0}& \mathbf{0} &\dots &\mathbf{E}_K\\
\end{bmatrix},
\end{equation}
resulting in the following simple form:
\begin{equation}\label{eq:Hop_mat}M=
\begin{pmatrix}
\boldsymbol{\omega}+2\Re\mathbf{D} & -i\mathbf{\Gamma} & -2\Re\mathbf{D} & -i\mathbf{\Gamma}\\
i\mathbf{\Gamma}^\dagger & \mathbf{\Xi} & -i\mathbf{\Gamma}^\dagger &\mathbf{\tilde{0}}\\
2\Re\mathbf{D} & -i\mathbf{\Gamma} & -\boldsymbol{\omega}-2\Re\mathbf{D} & -i\mathbf{\Gamma}\\
-i\mathbf{\Gamma}^\dagger &\mathbf{\tilde{0}}  & i\mathbf{\Gamma}^\dagger &-\mathbf{\Xi}\\
\end{pmatrix},
\end{equation}
where $\mathbf{\tilde{0}}$ is the zero-filled $3M\cdot K\times 3M\cdot K$ matrix.

We notice that the excitonic ($(|\beta|^2)$) and photonic ($(|\alpha|^2)$) single polariton fractions can be calculated from the normalized eigenvectors of the Hopfield matrix, expressed as follows:
\begin{equation}
    \ket{v}=\begin{pmatrix}
v_1 \\
\vdots \\
v_N\\
\vdots\\
v_{N+3M\cdot K}\\
\vdots\\
v_{2N+3M\cdot K}\\
\vdots\\
v_{2N+6M\cdot K}
\end{pmatrix} \, ,
\end{equation}
from which we define:
\begin{equation}
    |\alpha|^2= \mathlarger{\sum}_{i=1}^{N}|v_i|^2+\mathlarger{\sum}_{i=N+3M\cdot K+1}^{2N+3M\cdot K}|v_i|^2
\end{equation}
\begin{equation}
    |\beta|^2= \mathlarger{\sum}_{i=N+1}^{N+3M\cdot K}|v_i|^2+\mathlarger{\sum}_{i=2N+3M\cdot K+1}^{2N+6M\cdot K}|v_i|^2 \, .
\end{equation}
The latter accurately reproduce the single exciton and single photon fractions of the polariton eigenstates only when RWA can be assumed \cite{Quattropani1986,Ciuti2005}.

\end{document}